\documentclass[reprint, aps, prx, twocolumn,
superscriptaddress,
preprintnumbers,
amsmath,amssymb, footinbib]{revtex4-2}
\usepackage[english]{babel}
\usepackage{graphicx,dsfont}
\usepackage{color}
\usepackage{blindtext}
\usepackage{comment}

\usepackage{ulem}
\usepackage{tabularx}
\usepackage[%
colorlinks=true,
pdfborder={0 0 0},
allcolors = blue
]{hyperref}

\begin{document}
	
	\title{Current-induced self-organisation of mixed superconducting states}
	
	\author{X. S. Brems}
	\email{xaver.brems@frm2.tum.de}
	\affiliation{Heinz Maier-Leibnitz Zentrum (MLZ), Technische Universit\"at M\"unchen, D-85748 Garching, Germany}
	
	\author{S. M\"uhlbauer}
	\email{Sebastian.muehlbauer@frm2.tum.de}
	\affiliation{Heinz Maier-Leibnitz Zentrum (MLZ), Technische Universit\"at M\"unchen, D-85748 Garching, Germany}

	\author{W. Y. C\'ordoba-Camacho}
	\email{wilmercordoba@df.ufpe.br}
	\affiliation{Departamento de F\'isica, Universidade Federal de Pernambuco, Av. Prof. Aníbal Fernandes, s/n, 50740-560, Recife - PE, Brazil}
	
	\author{A.A. Shanenko}
	\email{ashanenko@hse.ru}
	\affiliation{National Research University Higher School of Economics, 101000, Moscow, Russia}
	\affiliation{Departamento de F\'isica, Universidade Federal de Pernambuco, Av. Prof. Aníbal Fernandes, s/n, 50740-560, Recife - PE, Brazil}
	
	\author{A. Vagov}
	\email{alexei.vagov@uni-bayreuth.de}
	\affiliation{Institut of Theoretical Physics Physics III, University of Bayreuth, Bayreuth 95440, Germany}
	
	\author{R. Cubitt}
	\email{cubit@ill.fr}
	\affiliation{Institut Laue-Langevin, 156X, 38042 Grenoble, France}
	
	\date{\today}

	\begin{abstract}
		
		Small-angle neutron scattering is used in combination with transport measurements to investigate the current-induced effects on the morphology of the intermediate mixed state domains in the intertype superconductor niobium. We report the robust self-organisation of the vortex lattice domains to elongated parallel stripes perpendicular to the applied current in a steady-state. The experimental results for the formation of the superstructure are supported by theoretical calculations, which highlight important details of the vortex matter evolution. The investigation demonstrates a mechanism of a spontaneous pattern formation that is closely related to the universal physics governing the intermediate mixed state in low-$\kappa$ superconductors.

	\end{abstract}
	
	\maketitle

	\section{Introduction}	
	Independent of their microscopic nature superconductors (SC) are usually categorized via their response to an external magnetic field. Materials only exhibiting complete flux expulsion (Meissner state) are classified as type I, whereas materials, showing the penetration of an array of supercurrent vortices in the mixed state, are referred to as type II \cite{Brandt_1995_flux}. For SC with a Ginzburg-Landau parameter close to $\kappa \approx \kappa_{0}\,(\kappa_{0} = 1/\sqrt{2})$, broadly referred to as intertype (IT) SC, this standard categorization breaks down \cite{Brandt2011}. A broad range of exotic magnetic flux patterns in the $\kappa$-$T$ plane emerges in vicinity of the Bogomolnyi point ($\kappa_{0},T_c$) due to the infinite degeneracy of the superconducting condensate, incompatible with the type I/II dichotomy \cite{vagov_universal_2020}. The different exotic states encountered in the IT regime, more specifically in its lower part below the line of the zero surface energy of the normal-superconducting interface, can not be solely explained by a non-monotonous vortex interaction of the two-body type, but rather need an interaction potential showing many-body characteristics \cite{wolf_vortex_2017}.
	
	The intermediate mixed state (IMS), the microscopic coexistence of complete magnetic flux expulsion (Meissner state) and the penetration of an array of supercurrent vortices (mixed state), is one of the most prominent examples of IT behavior in SC and has been studied extensively in several materials \cite{Ge2014_zrb12, Backs2019, Reimann2017, Reimann2015, Muehlbauer2009, Laver2006}. While sharing common features with the intermediate state (IS) of type I SC, where Meissner regions coexist with normal state domains, the IMS, in contrast to the IS, can not be solely explained by the effect of a non-zero demagnetization factor. It can only exist in the presence of a partially attractive interaction between vortices, which results in an equilibrium vortex distance $a_{0}$ \cite{Brandt2011}. Bitter decoration (imaging of vortices via Fe particle decoration \cite{Essman1967}) of pure Nb samples shows both laminar and tubular structure of the IMS \cite{Brandt2011} as also seen in the IS of type I SC \cite{Hoberg2008}. The behaviour of the IMS has been a subject of several theoretical and experimental studies in recent years \cite{Ge2014_zrb12, wolf_vortex_2017, vagov_universal_2020, Reimann2015, Reimann2017, Backs2019, Laver2006, Muehlbauer2009} where the temperature-field phase diagram in samples of different purity was mostly explored. Despite these efforts, properties of the IMS in IT SC are far from being fully understood.
	
	This work reports results of combined small-angle neutron scattering (SANS) and transport measurements of vortex clusters in a single-crystal bulk sample of the archetypal IT superconductor Nb. We investigate the changes in the IMS vortex configurations induced by an applied current demonstrating, that the current gives rise to a spontaneously emerging superstructure of parallel vortex stripes. It is well known that a current applied to a type II superconductor in the mixed state creates the Lorentz force acting on the vortices. It is balanced by the drag force resulting in a constant vortex velocity. In a typical type II superconductor these forces act similarly on all vortices, which then move with almost equal velocities and thus preserve their original Abrikosov lattice arrangement. 
	
	In contrast, our results reveal a totally different scenario for the vortex matter in the IMS. The initial configuration of isotropically distributed vortex clusters quickly rearranges itself by elongating in the direction perpendicular to the current flow. It eventually reaches a steady state of parallel vortex stripes. This state is robust - it is independent of the initial configuration and is conserved, when the current is rapidly switched off. The rearrangement dynamics suggests, that here - unlike in type II superconductors - the Lorentz and drag forces act differently on different vortices leading to a considerable vortex velocity dispersion at the initial evolution stage. 
	
	We argue that the appearance of the stripe superstructure is a generic phenomenon closely related to the physics of the IMS and the infinite degeneracy of the Bogomolnyi point. To demonstrate this we perform numerical simulations using the time-dependent Ginzburg-Landau (TDGL) model with two components, which is the simplest approach capturing essential qualitative characteristics of the stationary IMS \cite{da_silva_giant_2015}. The numerical simulations reveal details of the cluster elongation and its relation with the asymmetric current distribution and the vortex velocity dispersion. The experimental results and theoretical analysis demonstrate, that we are dealing with a remarkable example of the dynamical pattern formation. This places IT SC in line with a big group of systems where such self-organized phenomena take place, see, e.g., the Rayleigh-B\'{e}nard convection or Turing reaction-diffusion patterns in chemical reactions and biological systems \cite{Cross1993,Rabinovich2000,Pismen2006,Hoyle2006}.
	\section{Experimental Setup}
	\label{sec:expsetup}
\begin{figure*}
		\centering
		\includegraphics[width = \textwidth]{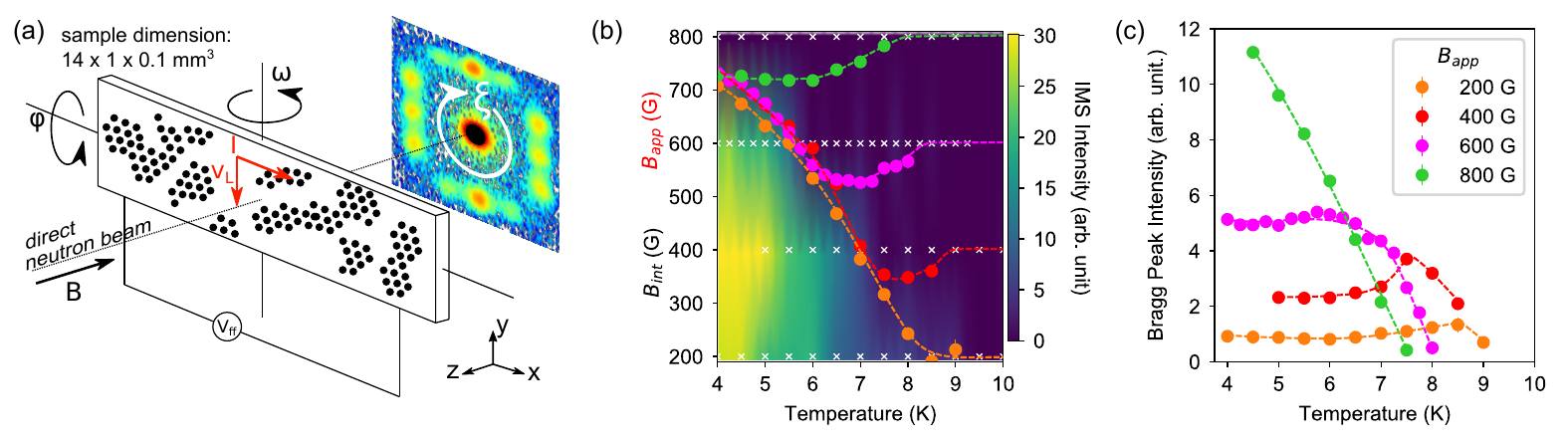}
		\caption{(a) Geometry of the SANS experiment (b) Average internal magnetic field $B_{int}$ and the colorplot of the IMS intensity. White crosses mark the applied magnetic field $B_{app}$ and temperature $T$. (c) Integrated intensities derived from the first order Bragg peaks. The dashed lines are guides to the eye.}
		\label{fig:fig1}
	\end{figure*}

	For our study, we used a combined transport measurement and small-angle neutron scattering (SANS) setup, consisting of a low temperature cryostat mounted inside an electromagnet, installed on the SANS diffractometer D33 at the Institut Laue-Langevin \cite{Dewhurst2016}. A schematic drawing of the measurement setup can be seen in Fig. \ref{fig:fig1} (a) with the orientation of the sample, current $I$ and applied magnetic field $B_{app}$. The sample stick was equipped with normal conducting copper current leads in the upper part and  NbSn superconducting current leads, spot welded to the sample  in the lower part, enabling high currents with minimized ohmic heating effects in the vicinity of the sample. For experiments dealing with the effect of current on the IMS, Helium was allowed to condense in the sample space at $T = 4\,\text{K}$, completely covering the sample in order to efficiently remove any heat created by vortex flow. Voltage contacts with a distance of $d\approx 8 \,\text{mm}$ required to record the characteristic I-V curves were made using silver conducting paint.
	
	We used a thin strip of a single crystal Nb sample ($14 \times 1 \times 0.1\,\text{mm}^3$) in our study. As indicated in Fig. \ref{fig:fig1} (a), the large face of the sample was aligned perpendicular to the magnetic field direction, which results in a large demagnetizing factor. Small cadmium sheets were used to mask the current and voltage leads. The current $I$ was applied perpendicular to the magnetic field $B$ along the sample. The magnetic field was aligned parallel to the direction of the incident neutron beam. Both cryostat and magnetic field could be rocked by the angles $\phi$ around a horizontal axis and $\omega$ around a vertical axis with both axes perpendicular to the neutron beam.
	
	For the SANS measurements, the collimation was set to 12.8\,m with a sample aperture of $\approx 5 \times 2\,\text{mm}^{2}$. The scattered neutrons were detected using a position-sensitive 2D detector placed 12.8\,m behind the sample. A medium resolution setup using a neutron wavelength $\lambda = 10\,\text{\AA}$ and square source aperture with $30 \times 30\,\text{mm}^{2}$ cross section was used for mapping the IMS phase diagram. For experiments dealing with the effect of current on the IMS we used a high resolution setup with neutron wavelength $\lambda = 14\,\text{\AA}$ and a round source aperture with $d = 20\,\text{mm}$ diameter. The full width half maximum wavelength spread was $\Delta \lambda / \lambda = 10 \% $ for both setups.
	
	The niobium single crystal sample was prepared by spark erosion. It was cut from a Nb single crystal obtained from Heraeus previously used in other experiments on the IMS \cite{Backs2019,Reimann2017}. The sample was left untreated after spark erosion cutting, since there are indications, that additional surface treatments (e.g. by means of electropolishing) increases the critical current \cite{Jones1966}. The demagnetization factor is $D=0.87$ ($B\perp$ to large sample face). From I-V measurements at room temperature and just above the transition temperature ($T_{S} = 9.5\,\text{K}$) a normal state resistivity of $\rho_{n}(9.5) = (3.7 \pm 0.2) \times 10^{-10} \,\Omega \, \text{m}$ and Residual Resistivity Ratio of $RRR \approx 390$ were deduced.

	\section{Results}
	\subsection{Stationary IMS}
	We first focus on the stationary IMS without applied current.
	Figure \ref{fig:fig1} (a) gives an example of a typical 2D SANS detector image of the scattering intensity profile, where the direct beam is excluded (central black circle).		
	The image shows the first order Bragg peaks stemming from the ordered flux line lattice inside the mixed state domains and the IMS scattering around the blacked out direct beam resulting from the magnetic contrast between Meissner state domains and mixed state domains. We observe 12 Bragg spots resulting from two isosceles domains of Nb ($\alpha \approx \beta \approx 65 ^\circ, \gamma \approx50^\circ$) as reported previously \cite{Muehlbauer2009}.
	Figure \ref{fig:fig1} (b) shows the temperature dependence of the average internal magnetic field $B_{int}$ for a few selected values of the applied field $B_{app}$.
	The value of $B_{int}$ is 	obtained using the relation
	\begin{equation}
	\centering
	B_{int} = \frac{\phi_{0} \sqrt{3}q_{VL}^2}{8\pi^2}, 
	\label{eq:Bint}
	\end{equation}
	where $\phi_{0}$ is the elementary flux quantum and $q_{VL}$ is the average $|\mathbf{q}|$-value of the first order Bragg peaks \footnote{In this equation we used the approximation of a perfect hexagonal lattice, since we weren't able to resolve the exact position of the first order Bragg peaks for low applied magnetic fields and temperatures inside the IMS regime. Therefore it was not possible to extract the exact reciprocal unit cell, which has been shown to change as a function of applied field $B_{app}$ and temperature $T$ \cite{Muehlbauer2009}. Equation (\ref{eq:Bint}) still gives a good approximation of the internal field $B_{int}$, since the observed angles at high applied magnetic fields and low temperatures are close to $60^{\circ}$ and we only assume slight deviations for lower applied magnetic fields and high temperatures.}.
	The color density map in Fig. \ref{fig:fig1} (b) represents the IMS scattering obtained by summing the intensity around the direct beam [between the white circles in Fig. \ref{fig:fig2} (a)].
	The data were collected during warming (W) after field cooling (FC) in an applied magnetic field $B_{app}$ from $T = 10\,\text{K}$ to $T = 4\,\text{K}$ and are corrected for a high temperature background ($T = 9.2\,\text{K}$).
	In the data analysis we assume that the diamagnetic response after FC is vanishingly small using the fact that other samples obtained from the same single crystal demonstrate a negligible diamagnetic contribution in FC magnetization measurements \cite{Backs2019}.
	
	At low temperatures, the internal field $B_{int}$ in Fig. \ref{fig:fig1} (b)  follows the same universal temperature dependence $B_{int}^\ast(T)$ for all values of the applied field $B_{app} < 800\,$G. This behavior is known to be a hallmark of the IMS \cite{Backs2019}. When $T$ increases further, $B_{int}(T)$ departs from $B_{int}^\ast(T)$ eventually approaching the corresponding value of $B_{app}$ in a way typical for the conventional mixed state.
	The crossover between the IMS and the conventional mixed state is also observed in the colour density plot in Fig. \ref{fig:fig1} (b) representing the summed IMS scattering intensity around the direct beam, called IMS intensity.
	Starting at low temperatures, the IMS intensity  peaks at an applied magnetic field of $B_{app} = 400\,\text{G}$. It decreases for increasing and decreasing magnetic fields. For a given applied magnetic field $B_{app}$, the IMS intensity decreases with rising temperature and vanishes once the internal magnetic field $B_{int}$ leaves the common temperature dependence. Figure \ref{fig:fig1} (c) shows the integrated intensities derived from the first order Bragg peaks. In agreement with \cite{Backs2019}, we see a linear increase with falling temperature (typical of the vortex lattice in the standard mixed state) and observe a downward bent curve (600\,G) and an additional kink (400\,G and 200\,G) most pronounced at $B_{app} = 400\,$G, indicating the transition to the IMS for applied magnetic fields $B_{app}<800\,\text{G}$. Inside the IMS the integrated intensity doesn't decrease with decreasing temperatures, as would be expected when approaching the Meissner state with vanishing internal field. We attribute this to considerable pinning effects. Note however, that these pinning effects don't hinder a microscopic rearrangement of the vortex lattice, when entering the IMS regime as seen from the change in internal magnetic field $B_{int}$ with $T$ shown in Fig. \ref{fig:fig1} (b) and reported elsewhere \cite{Backs2019}. The temperature of the observed kink / the start of the downward bent part agrees sufficiently well with the temperature of deviation from $B_{int}^\ast(T)$ shown in Fig. \ref{fig:fig1} (b). In summary we established the range of the IMS in good agreement with literature \cite{Backs2019}.
	\subsection{Current induced changes in the IMS}
	\begin{figure}
		\centering
		\includegraphics[width = \columnwidth]{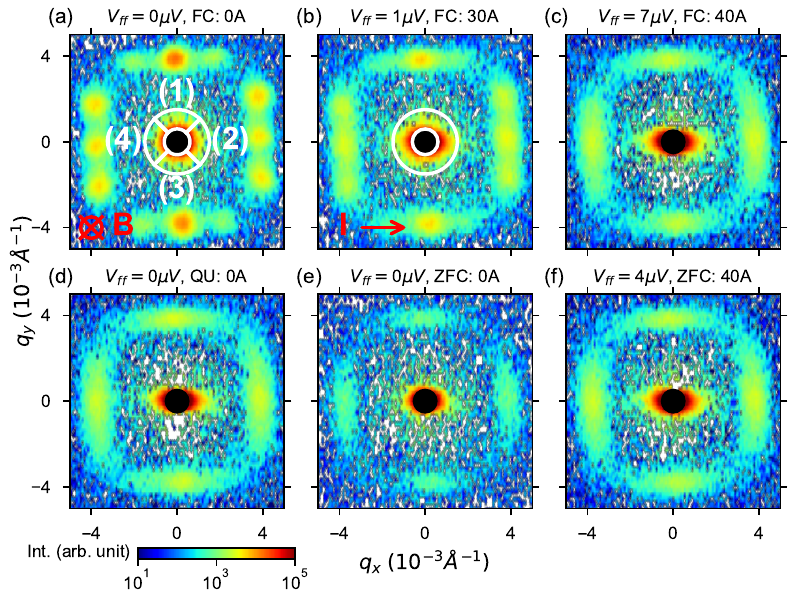}
		\caption{SANS images from applied current measurements in a magnetic field of $B_{app} = 500$ G at $T = 4$ K: (a) - (c) FC approach with incremental current ramp from $I = 0$ A up to $I = 40$ A. (d) Subsequent current quench (QU) to $I = 0$ A. (e) ZFC approach without a current and (f) subsequent current ramp to $I = 40$ A. The measured flux flow voltage $V_{ff}$ is shown above each panel.}
		\label{fig:fig2}
	\end{figure}
	We will now investigate changes in the IMS induced by the applied current $I$. Previous to the neutron experiment we identified the regime of the flux flow by measuring the I-V curves of the sample for different values of $B_{app}$ and $T$. The results are shown in Fig. \ref{fig:sup1} in Appendix \ref{app:Transport}.
	
	Figure \ref{fig:fig2} shows examples of SANS images obtained with various applied currents in a magnetic field of $B_{app} = 500$ G at $T = 4$ K under liquid He. Shown is the sum over rocking angles in the interval $\phi, \omega \in [-4^\circ,5^\circ]$. The current is applied in $x$-direction perpendicular to the magnetic field, as indicated in panels (a) and (b). The flux flow voltage $V_{ff}$ is shown above the panels. All the data were corrected for a zero field cooled (ZFC) background at $T = 4\,\text{K}$.
	
	Figures \ref{fig:fig2} (a - d) show measurements following a FC protocol: The sample was cooled in an applied magnetic field of $B_{app}=500\,$G with no applied current from $T = 10\,\text{K}$ to $T=4\,\text{K}$. We then performed rocking scans without applied current [Fig. \ref{fig:fig2} (a)] and after applying an external current in incremental steps up to $I = 30\,\text{A}$ [Fig. \ref{fig:fig2} (b)] and after further increasing to $I = 40\,\text{A}$ [Fig. \ref{fig:fig2} (c)]. While still being in an applied magnetic field $B_{app}  =500\,\text{G}$ and cold at $T=4\,\text{K}$ we quenched the current to $I = 0\,\text{A}$ and subsequently performed another set of rocking scans shown in Fig. \ref{fig:fig2} (d).
	
	We also employ a different protocol where the sample is first cooled from $T = 10\,\text{K}$ to $T = 4\,\text{K}$ in the zero field (zero-field cooling - ZFC) and then the field is ramped up to  $B_{app} = 500\,\text{G}$. The SANS image obtained after this protocol for $I=0$ is shown in Fig. \ref{fig:fig2} (e) while Fig. \ref{fig:fig2} (f) illustrates the changes after the current is increased to $I = 40\,\text{A}$.
	
	We first consider the Bragg peaks stemming from the vortex lattice in the mixed state domains.
	For the FC protocol at $I = 0\,\text{A}$ [Fig. \ref{fig:fig2} (a)] and up to $I = 20$\,A (not shown) both the position and intensity of the Bragg peaks from two well ordered isosceles vortex lattice domains remain unchanged. When the current passes the critical value $I\gtrsim 30$\,A and the vortices start to move resulting in $V_{ff} \neq 0$ the Bragg peaks get smeared out in the azimuthal direction and decrease in intensity  [Figs. \ref{fig:fig2} (b,c)] due to the lattice disorder induced by the motion. The disorder is retained after the current is quenched to $I = 0\,\text{A}$ [Fig. \ref{fig:fig2} (d)].
	
	The ZFC - field ramp protocol does not create a well ordered vortex lattice as is indicated by the smeared Bragg peaks in Fig. \ref{fig:fig2} (e).
	An external current of $I = 40\,\text{A}$ [Fig. \ref{fig:fig2} (f)] restored the equivalent FC case [Fig. \ref{fig:fig2} (c)] showing an ordered vortex lattice with azimuthally smeared out Bragg peaks. The Bragg peaks obtained after the FC and ZFC - field ramp protocol are practically indistinghuishable in the 2D SANS image.
	A further detailed analysis of the vortex lattice Bragg peaks can be found in Appendix \ref{app:Fluxbending} and Appendix \ref{app:rad_width} (Fig. \ref{fig:sup2} and Fig. \ref{fig:sup3}).
	
	Details of the IMS structure are reflected by the scattering in the vicinity of the direct beam -  between the white circles in Fig. \ref{fig:fig2} (a).
	For the FC protocol and $I = 0\,\text{A}$ [Fig. \ref{fig:fig2} (a)] up to $I=20$\,A (not shown) the IMS scattering is isotropic.
	However, when  $I \geq 30$A  [Figs. \ref{fig:fig2} (b,c)] one observes a notable anisotropy in the scattering pattern which becomes elongated in the horizontal $x$-direction. The anisotropy is preserved after a rapid current quench to $I = 0\,\text{A}$ [Fig. \ref{fig:fig2} (d)]. After the ZFC - field ramp protocol the IMS scattering also demonstrates a slight anisotropy in horizontal direction [Fig. \ref{fig:fig2} (e)].  A subsequent current ramp to $I = 40\,\text{A}$ results in the same IMS scattering pattern as obtained after the FC protocol [cf. Fig. \ref{fig:fig2} (c) and (c)] showing a clear anisotropic IMS scattering around the direct beam.
	
\begin{figure}
		\centering
		\includegraphics[width = \columnwidth]{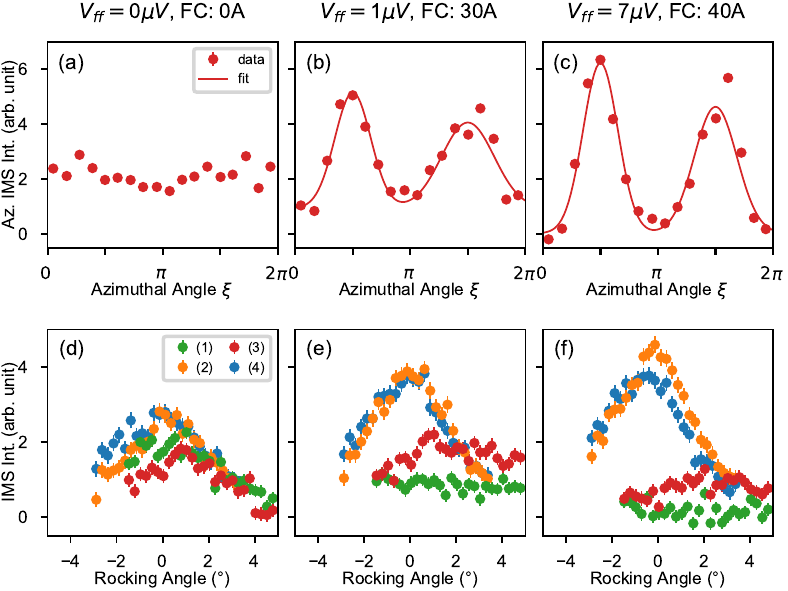}
		\caption{Azimuthal averages [(a)-(c)] and rocking curves [(d)-(f)]  of the IMS scattering in current experiments in a magnetic field of $B_{app} = 500\,$G at $T = 4$\,K. The solid line is a fit of two Gaussians. Shown are rocking scans around a horizontal axis $\phi$ [curves (1) + (3)] and vertical axis $\omega$ [curves (2) + (4)].  The sectors for the azimuthal averaging and rocking scans are depicted in Fig. \ref{fig:fig2} (b) and Fig. \ref{fig:fig2} (a), respectively. The current $I$ and the voltage $V_{ff}$ are shown above each column.}
		\label{fig:fig3}
	\end{figure}

	Further details of the IMS scattering are presented in Figs. \ref{fig:fig3} (a - c) that plot the radially averaged IMS scattering intensity as a function of the azimuthal angle $\xi$
	The azimuthal angle and the sector to evaluate the IMS scattering are shown in the SANS image of Fig.\,\ref{fig:fig1} (a) and in Fig.\,\ref{fig:fig2}  (b) between white circles, respectively \footnote{To obtain the angle dependence shown in Figs. \ref{fig:fig3} (a - c) we take the intensity in Figs. \ref{fig:fig2} (a - c), respectively,  represent it as function of the radial distance from the beam center and the azimuthal angle $\xi$ and then integrate the intensity over the radial distance inside the sector of interest}.
	The detector pixel size and the small scattering angles limit the azimuthal bin size to $\Delta \xi = 20^\circ$. The corresponding values of $V_{ff}$ and $I$ are shown above each panel. 
	
	At $I = 0$ [Fig. \ref{fig:fig3} (a)] up to $I=20\,$A (not shown) the IMS scattering is isotropic being independent of the azimuthal angle $\xi$. At $I\ge30\,$A the scattering shown in Figs. \ref{fig:fig3} (b,c) is angle-dependent
	having its maxima at $\xi = \pi/2$ and $3\pi/2$ (in the horizontal $x$-direction) and minima at at $\xi = 0, \pi$ and $2\pi$ (in the vertical $y$-direction). When the current increases to  $I = 40\,$A the difference between the maximal and minimal values increases [see Fig. \ref{fig:fig3} (c)]. 
	
	Finally, Figs. \ref{fig:fig3} (d - e) show the rocking curves of the IMS scattering inside the sectors $1-4$ shown in Fig.\,\ref{fig:fig2} (a) \footnote{To obtain the rocking curves we take the rocking scan intensity (not summed over angles $\phi$ and $\omega$ as is done in Figs. \ref{fig:fig2}), and sum the intensity inside the sectors of interest for each rocking angle $\phi$ and $\omega$. The magnetic field was aligned via the standard method of using the Bragg spots of the vortex lattice. Rocking angles $\phi<-2^\circ$ were not achievable due to the size of the opening window of the electromagnet.}.
	The summed scattering intensities corresponding to sectors 1 and 3 are represented by green and red circles, respectively, and plotted as functions of the rocking angle $\phi$ around the horizontal axis. The summed scattering intensities inside sectors 2 and 4 are represented by orange and blue circles, respectively, and plotted as functions of the rocking angle $\omega$ around the vertical axis. When the current is absent, $I = 0\,\text{A}$, the IMS intensity dependence on the rocking angle is qualitatively similar for all four sectors [Fig. \ref{fig:fig3} (d)]. However, at $I = 30\,\text{A}$ Fig. \ref{fig:fig3} (e) reveals a clear difference between the rocking scans of the horizontal sectors (1 and 3) and rocking scans of the vertical sectors (2 and 4). The intensity in sectors 2 and 4 increases and develops a sharper angle dependence, whereas the rocking curves of sectors 1 and 3 flatten. This effect is getting more pronounced when the current increases [Fig. \ref{fig:fig3} (f)].
	
	\section{Discussion}
In the absence of transport current, the characteristics of the IMS such as depicted in Figs. \ref{fig:fig1} (b,c) are in agreement with earlier experiments \cite{Backs2019}. 
The hallmark of the IMS, a $B_{app}$ independent, but $T$ dependent internal magnetic field $B_{int}$ and the presence of very small angle scattering inside the IMS regime is cearly observed. 

In our study we see, that the IMS still persists in the state of flux flow, as evidenced by I-V-characteristics in different applied magnetic fields (see Fig. \ref{fig:sup1} in Appendix \ref{app:Transport}) and the neutron data. Figure \ref{fig:fig2} clearly demonstrates, that the hallmarks of the IMS - very small angle scattering and the constant internal field $B_{int}$ - are preserved in the state of vortex movement.

Moreover, applied current gives rise to an elongation of the IMS domains, that coincides with the onset of flux flow. The elongation is characteristic of a sheet pattern orthogonal to the applied current and parallel to the movement of flux lines.
The elongation is manifested by the transition from isotropic to anisotropic IMS scattering parallel to the applied current in $x$ direction at the onset of flux flow at $I_{c} \approx 30$\,A, as seen in Fig. \ref{fig:fig2} (b) and Fig. \ref{fig:fig3} (b). 
The anisotropy in the IMS scattering [Fig. \ref{fig:fig2} (c) and Fig. \ref{fig:fig3} (c)] is more pronounced with increasing current up to the maximal applied current $I = 40$\,A.

Furthermore, the rocking curve of the IMS scattering with respect to the vertical axis $\omega$ becomes more pronounced, while the rocking curve around a horizontal axis in $\phi$ flattens with increasing current [Figs. \ref{fig:fig3} (d - f)]. This behaviour is expected for scattering from a sheet-like superstructure.

At the same time, the current induces an overall increase of disorder of the vortex lattice. This fact is evident from the broadening of the vertical and horizontal Bragg spots of the vortex lattice with increasing current, seen from the increased azimuthal smearing (Fig. \ref{fig:fig2}), the radial width (Fig. \ref{fig:sup3} in Appendix \ref{app:rad_width}) and the rocking curve width (Fig. \ref{fig:sup2} in Appendix \ref{app:Fluxbending}) of the Bragg peaks, as well as from the corresponding loss of intensity. The latter cannot be attributed to ohmic heating, since a temperature increase would also lead to a decrease in the total IMS intensity [Fig. \ref{fig:fig1} (b)], not seen in Figs. \ref{fig:fig3} (a - c).

A careful study of the rocking curve shape of the vortex lattice Bragg peaks as presented in Appendix \ref{app:Fluxbending} allows for an estimation of the vortex bending due to the magnetic field induced by the applied current denoted magnetic self field. The applied current splits into a bulk component $I_{bulk}$ and a non-dissipative surface component $I_{surf}$. Assuming the bulk current equals the applied current would result in a flat top shape of the rocking curves of vertical Bragg spots with a broadening of almost $30\,^\circ$ due to the additional magnetic self field with a maximal value at the surface of $B_{surf} \approx 250\,$G. In contrast, our measurements only show an additional broadening of $\Delta \sigma \approx 0.5 - 0.8 ^\circ$  at $I = 40\,$A (rocking curves of the vertical Bragg spots with respect to the rocking curves of the horizontal Bragg spots). Our calculations therefore suggest a dissipationless critical current of $I_{c} =  36$\,A flowing on the sample surface and the excess current of $I_{bulk} = 4\,$A penetrating the bulk showing good agreement with the measured rocking curve. This splitting into bulk current and dissipationless surface current has also been shown previously \cite{Hocquet1992,Pautrat2003, Goupil2000} and is further strengthened by the shape of the I-V-curve, showing the absence of a discontinuity at the critical current $I_{c}$ (Fig. \ref{fig:sup1} in Appendix \ref{app:Transport}) in agreement with literature \cite{Hocquet1992}.

The current-driven IMS achieves a steady state, which represents a key result of this study. The steady state is robust, remaining after the current is abruptly quenched to zero. This is illustrated in Fig. \ref{fig:fig2} that demonstrates no visible changes in the small-angle scattering pattern. Both, the elongated IMS superstructure and the disorder in the vortex lattice are retained, when the current is off, with the exception of the additional broadening associated with the current-induced field. Furthermore, the transport current suppresses the hysteretic behaviour of the IMS pattern with respect to FC-ZFC protocols. When the current is absent the ZFC protocol does not produce a well ordered lattice \cite{Backs2019}. A slight anisotropy in the $\mathbf{q}_{x}$ direction observed after applying the ZFC protocol is explained by flux entering the sample overcoming the Bean barrier \cite{Bean1964} preferentially via the short sample dimension and therefore forming elongated IMS domains in the vertical $y$ direction in Fig. \ref{fig:fig1} (a). However, a subsequent current ramp to $I = 40$\,A drives the system to the state, which is achieved after applying the FC protocol for the same current. This fact is evident in both the Bragg peak and very small-angle IMS scattering shown in Fig. \ref{fig:fig2}.

Our experimental results are compared with theoretical calculations. We use a model of two coupled Ginzburg-Landau (GL) equations (see Appendix \ref{app:theory}) to simulate the time evolution of IMS vortex configurations. The validity of this model for single-band IT superconductors follows from the underlying physics of the IMS related to the self-duality of the BCS theory at the infinitely degenerate Bogomolnyi point ($T \to T_c$ and $\kappa \to \kappa_0$)  \cite{vagov_superconductivity_2016, vagov_universal_2020}. At  $T < T_c$ and $\kappa \neq \kappa_0$ the degeneracy is lifted creating IMS configurations. This mechanism is generic and qualitatively independent of the system details. Two coupled GL equations is one of the simplest models that captures key features of the stationary IMS in low-$\kappa$ materials \cite{da_silva_giant_2015}. It yields the same phase diagram of the IT domain as does the BCS theory with a single band  \cite{vagov_superconductivity_2016} and reproduces fine details of multi-vortex interactions  \cite{wolf_vortex_2017}. Here we extend this model to describe the current-driven evolution of the IMS. Our calculations do not account for the additional factors like anisotropy, pinning, sample geometry, and the stray field, which, although important for quantitative characteristics, do not affect the qualitative picture.   
\begin{figure}
	\centering
	\includegraphics[width = 1\columnwidth]{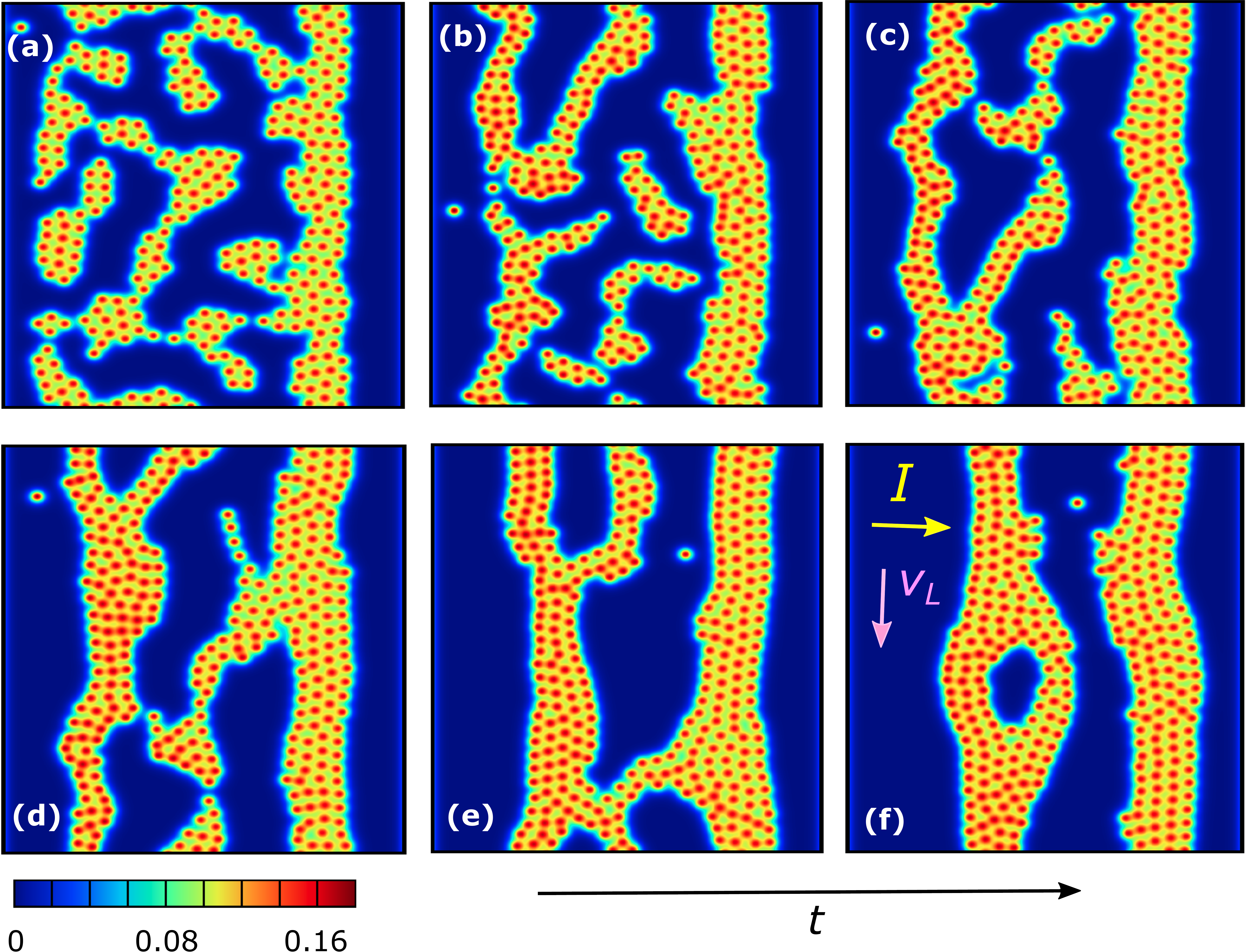}
	\caption{Numerical simulations of the IMS time evolution (the spatial distribution of the magnetic field is given). Panel (a) represents the initial configuration of the vortex clusters, panels (b)-(f) give snapshots at later times. The flowing current $I$ and the vortex motion with velocity $v_L$ are shown by arrows in panel (f). Details of the model and its microscopic parameters are described in Appendix \ref{app:theory}. }
	\label{fig:fig4}
\end{figure}
When the current in the sample exceeds the critical value $I>I_c$, vortices start moving in the perpendicular direction to the current flow due to the Lorentz force, and the IMS configuration changes. Figure \ref{fig:fig4} shows snapshots of the time evolution of the initial IMS with random vortex clusters in Fig. \ref{fig:fig4} (a). While moving, the clusters elongate in the direction parallel to their movement and merge. Finally, the time evolution creates a superstructure of vortex  stripes elongated perpendicularly to the average current. Results of the simulations are fully consistent with the experimental observations. The type of the vortex structure inside the clusters depends on a point in the IT phase diagram that corresponds to the superconducting material \cite{vagov_universal_2020}. When the material is close to type II ($\kappa \gtrsim \kappa_0$) vortices form a lattice, while in materials close to type I ($\kappa \lesssim \kappa_0$) the lattice melts becoming a liquid. However, for any initial IMS configuration, the current-driven evolution eventually yields qualitatively similar arrangements of vortex stripes.

\begin{figure}
	\centering
	\includegraphics[width = 0.7\columnwidth]{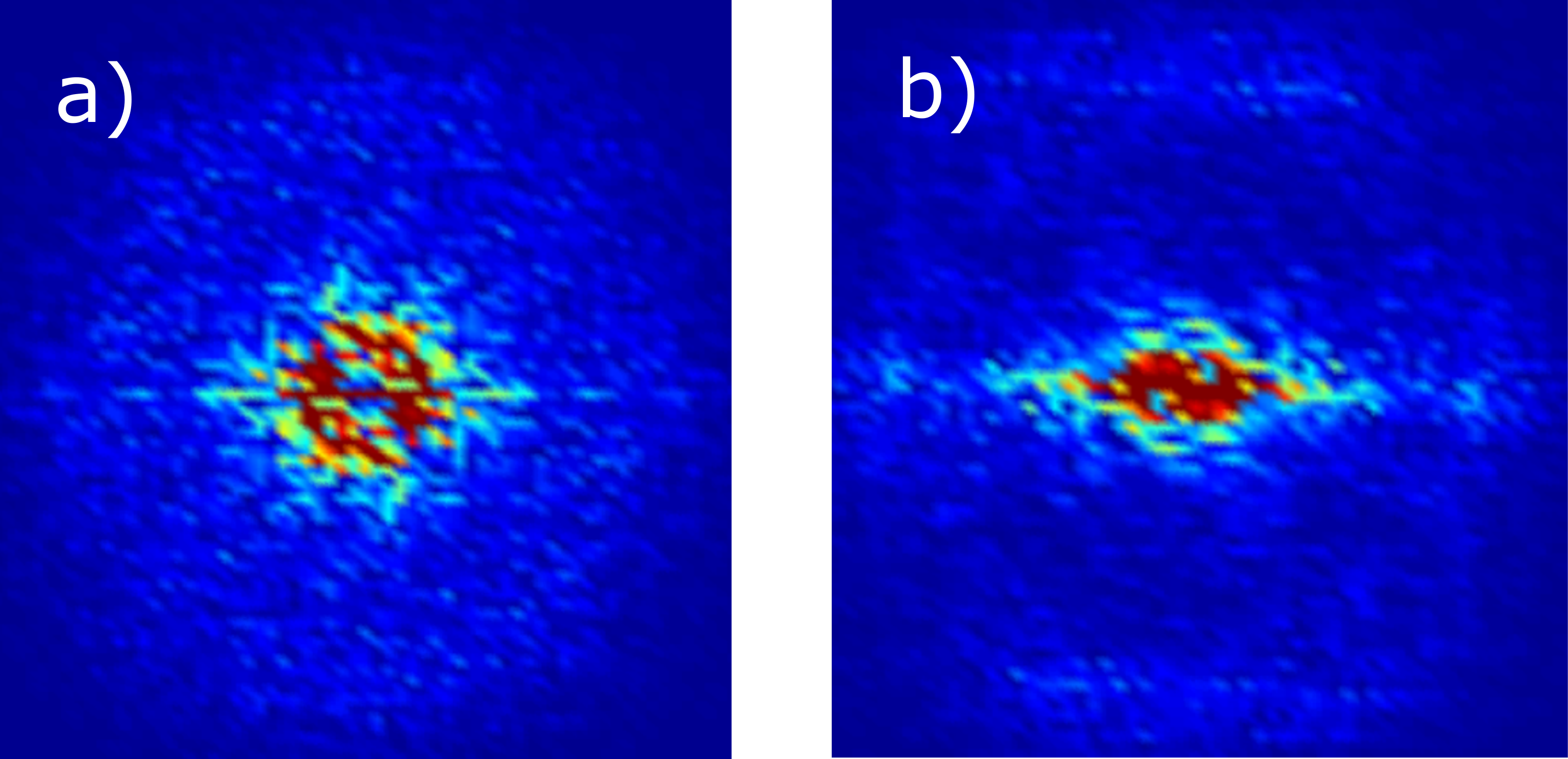}
	\caption{The Fourier transform of the numerically obtained field distribution (Panel a) at the start of the current-induced evolution shown in Fig. \ref{fig:fig4} (a), and (Panel b) when the stripes are formed as shown in Fig. \ref{fig:fig4} (f). Appearance of stripes is reflected in the elongated central (small angle) part of the Fourier transform.}
	\label{fig:fig5}
\end{figure}

The elongation of the vortex structures and subsequent formation of the stripe superstructure can also be traced in the Fourier transform (FT) of the spatial field profile, which corresponds directly to the measured intensity in a scattering experiment.
The FT calculated from the initial field configuration in Fig. \ref{fig:fig4} (a), obtained before the current is applied,  is shown in Fig. \ref{fig:fig5} (a). Correspondingly, Fig. \ref{fig:fig5} (b) shows the results of the FT for the stripe configuration in Fig. \ref{fig:fig4} (f).  One sees that the initially almost isotropic FT profile changes to the visibly elongated structure. This fully agrees with the SANS images in Fig. \ref{fig:fig2}. The absence of vortex lattice Bragg peaks in the FT is explained by the simplified numerical simulation neglecting the crystal anisotropy.

It is worth noting that several recent studies \cite{Dobrovolskiy2020, embon_2017} on type II superconducting films also demonstrated spontaneously emerging vortex configurations, elongated in the direction perpendicular to the current flow, when the applied current approached the pair-breaking limit. Those "vortex rivers" appear similar to quasi-1D vortex chains and enable an ultra-fast vortex motion inside them. In contrast, the stripes observed in our study are much wider and the rearrangement takes place at much lower currents only slightly exceeding the critical value of vortex movement.

The rearrangement process into the elongated IMS superstructure is a result of a combined action of three main factors: the current-induced Lorentz force, the drag force and the inter-vortex interaction potential. Their action on the IMS in IT superconductors demonstrates peculiar features, that distinguish these materials from conventional type II superconductors. In IT materials, the vortex core size is comparable with the inter-vortex distances resulting in spatially non-monotonic vortex interactions with sizable multi-vortex contributions \cite{da_silva_giant_2015,wolf_vortex_2017}. As a result, vortices necessarily clusterize and can no longer be viewed as isolated "elementary particles" of the mixed state \cite{vagov_universal_2020}. Consequently, the IMS is characterized more by the collective vortex behavior than by the properties of an individual vortex. The total action of the applied current thus depends on a configuration of vortex clusters. In principle, this can be seen already in type II superconductors where the current modifies the vortex–vortex interactions such that it starts to depend on the vortex orientation relative to the current flow \cite{kogan_2020}. The proximity to the degenerate Bogomolnyi point enhances this geometry dependence. The calculations show that the applied current flows primarily along boundaries between the mixed and Meissner state resulting in the Lorentz force that pulls that boundary in normal direction towards the Meissner phase (see Fig. \ref{fig:sup4} in Appendix \ref{app:elongation}). The asymmetry of the current flow creates a non-zero cumulative force which depends on the configuration size and shape. This results in the dispersion of the vortex velocities, which in turn distorts the configuration shape. Numerical calculations demonstrate that vortex structures, that are large and elongated in the motion direction, move faster  (see Appendix \ref{app:elongation} for details of the time evolution of vortex clusters of different shape and size). It is intuitively explained by the larger current density and thus larger Lorentz force at the boundary of a larger cluster. Combined with the drag force and the non-monotonic vortex interaction, that keeps the inter-vortex distance, this configuration-dependent Lorentz force further elongates vortex structures in the direction perpendicular to the applied current eventually creating a superstructure of stripes as shown in Fig. \ref{fig:fig4}.

	We finally discuss the limitations of our theoretical approach before we  highlight future perspectives of transport phenomena in the IMS to model percolation and non-trivial flow phenomena in two domain systems.
	Our model of two coupled TDGL equations does not take into account many factors such as pinning, anisotropy, and phonon-induced relaxations. It also uses a simplified model that assumes that the average current is injected through the system and then redistributed by the vortex configuration. In real samples the current in the inner part can only flow in regions with non-vanishing curl of $B$ according to Ampere's law. Transport current is therefore constrained to the mixed state, the surface and the interfaces between the Meissner and the mixed state domains. In turn, current cannot exist in the Meissner regions where the magnetic field is zero. In order to pass a current through the bulk of our sample, we therefore need a connected mixed state bridging the contacts. For our experiment, the volume fraction of the mixed state is estimated using $f_{MS} = V_{MS}/V =  B_{app} / B_{int}$ with the total sample volume $V$ and the mixed state volume $V_{MS}$. This estimation holds since we expect a negligible diamagnetic behaviour in our sample \cite{Backs2019} and therefore the applied magnetic field is equal to the internal magnetic field averaged over the whole sample volume. In our case this estimate  yields $f_{MS} \approx 70\,\%$  at the field $B_{app} = 500\,$G. This is above the percolation threshold of $\rho_{c} = 44 \,\%$ in 2D systems \cite{Zallen1970}. Therefore we can safely assume the connectivity of the mixed state domains of our sample takes place at the onset of the flux flow, where we made use of the quasi two dimensional character of the vortex lattice. This renders our simplified theoretical model as a valid description of our experimental findings.
	
	We note, however, that at lower magnetic fields and the associated lower volume filling of the IMS below the percolation threshold, one expects a much more complex situation. When the connectivity is broken, one needs to employ a fully inhomogeneous theoretical description for a finite sample that takes into account the influence of the stray fields outside the sample to capture the essential physics of the system. In this case two orthogonal flows restricted to a single phase of a two phase system might lead to interesting ordering phenomena. Further complications arise due to the influence of the surfaces of the sample, where (i) the vortices are nucleated and destroyed, (ii) pinning is significantly different as compared to the bulk and (iii) the influence of the dominant surface current is unclear. Besides the unknown balance of surface versus bulk pinning, it is also unclear whether the IMS structure is nucleated at the surface of the sample or forms as a steady state deep inside the bulk of the material.

	\section{Conclusion}
	This work studies the evolution of the IMS domains in the IT superconductor Niobium under the influence of an external transport current using a combined SANS and transport measurement technique.
	
	The study demonstrates a transition from isotropic to anisotropic IMS scattering, indicating, that the IMS rearranges itself into a stripe superstructure in the regime of flux flow. The stripe pattern is aligned perpendicular to the current direction along the motion of the vortices. A close examination of the rocking scans showed a splitting of the current into bulk and surface component with the latter being dominant. Most importantly, the absence of the hysteretic behaviour proves that the elongated superstructure is a steady state in the flux flow regime. Numerical simulations of the time evolution of IMS vortex configurations using a model of two coupled Ginzburg-Landau (GL) equations qualitatively reproduced our experimental results and revealed details of the cluster elongation.
	Our findings highlight the importance of the IMS as a model system for universal domain physics and demonstrate,  that  we  are  dealing  with  a  remarkable example of a self-organized pattern formation phenomenon.
	The current-induced movement of vortices in the IMS might act as a model system for the study of percolation and non-trivial (orthogonal) flow and self-ordering phenomena in two domain systems.

\begin{acknowledgments}
	We express our gratitude to F. Marchal and F. Lapeyre for their support with sample preparation and M. Bonnaud for support with the experiments. Further thanks are due to A. Backs for fruitful discussions and his contributions to this project in the early stage. 
	This work is based upon experiment EASY-568 \cite{D33_2020_01} and experiment 5-31-2748 \cite{D33_2020_09} performed at the instrument D33 at the Institut Laue Langevin (ILL), Grenoble, France.
\end{acknowledgments}

\appendix
\section{Transport Measurement}
\label{app:Transport}
We used a dedicated combined transport measurement and SANS setup presented in Sec. \ref{sec:expsetup} of the main text, which allowed us to study the I-V-characteristics of our sample prior to the experiment and also monitor the voltage response of our sample to an external current during the SANS experiment.
During pre-characterization we measured I-V-curves at a multitude of different magnetic fields $B_{app}$.
In contrast to the neutron experiment, we didn't follow a FC measurement protocol for the pre-characterization measurements, since changing magnetic field $B_{app}$ implies heating up the sample above the transition temperature and therefore boiling away the condensed He in the cryostat. The data points of the I-V-curves of the pre- characterization were recorded on an average frequency of $\approx 1/10\,\text{s}^{-1}$.

During the neutron experiment and the measurement of the I-V-curve the sample was FC.
The I-V-data points during the neutron experiment represent an average over the voltage recorded over the whole duration of a rocking scan, which leads to an acquisition time of  $\approx 2.5\,\text{h}$ per point.

\begin{figure}
	\centering
	\includegraphics[width = \columnwidth]{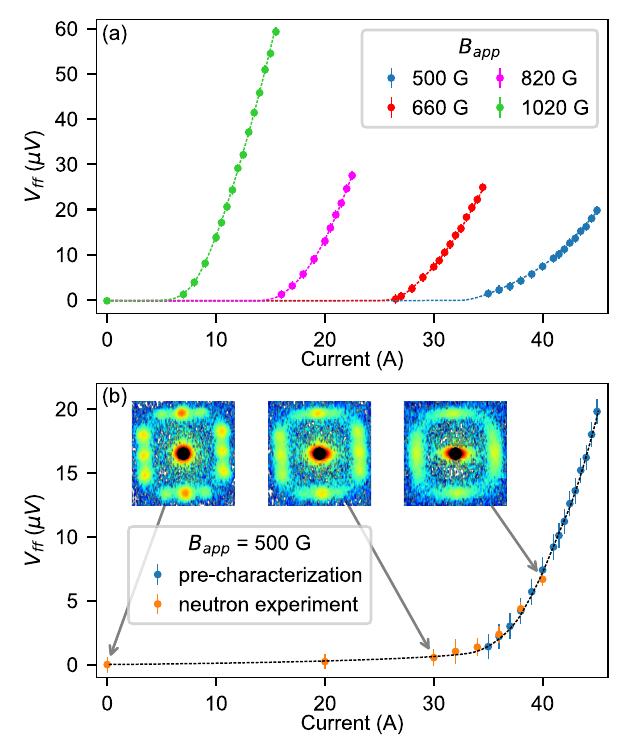}
	\caption{(a) Transport measurements prior to the neutron experiment at $T = 4$\,K as a function of different applied magnetic fields $B_{app}$.
		(b)	Comparing the I-V-curve measured prior to the experiment with the in-situ measurement in a magnetic field of $B_{app} =  500$\,G at $T = 4$\,K. The insets show the corresponding 2D detector image.  The dashed lines are guides to the eye. }
	\label{fig:sup1}
\end{figure}

Figure \ref{fig:sup1} summarizes the I-V-characteristic of our sample at $T = 4\,\text{K}$ under liquid He at different applied magnetic fields $B_{app}$.
Figure \ref{fig:sup1} (a) shows examples of I-V-curves at $T = 4\,\text{K}$ in different applied magnetic fields $B_{app}$ collected prior to the neutron experiment. As expected from literature \cite{Huebener1970}, we see a decrease in the critical current for flux flow with increasing applied magnetic field $B_{app}$.
Furthermore the slope of the I-V-curve related to the flux flow resistance $R_{ff}$ is increasing with increasing applied magnetic field. 

For a normal \mbox{type-II} SC, exhibiting an Abrikosov lattice covering the whole sample for applied fields $B_{app} >B_{c1}$, the flux flow resistance has been shown to be proportional to the applied field $B_{app}$ according to Eq. (\ref{eq:R_ff}) derived in the Bardeen-Stephen model with the upper critical magnetic field $B_{c2}$ and the normal state resistance $R_{n}$ \cite{Bardeen1965}.

\begin{equation}
\label{eq:R_ff}
R_{ff} \propto R_{n}\frac{B_{app}}{B_{c2}}.
\end{equation}

This should still hold for an IT SC in the pure mixed state. Inside the IMS regime, the sample splits into domains of mixed state and Meissner state. Assuming a homogeneous current distribution constraint to mixed state domains of the sample leads to a local decrease in the sample cross section and therefore an increased local resistance by a factor of $B_{int} / B_{app} = f_{MS}^{-1}$.  The measured voltage is an average over the whole sample and since only the mixed state contributes to the voltage build-up, we get an additional factor of $B_{app} / B_{int} = f_{MS}$, which cancels the factor resulting from the local cross section decrease. This however only holds, if we have a sample with zero magnetization.

Figure \ref{fig:sup1} (b) shows the I-V-curve measured prior to the neutron experiment (blue curve) and the I-V-curve derived from the voltage measurement during the neutron experiment (orange curve), both in an applied magnetic field of $B_{app} = 500\,\text{G}$. The 2D detector images corresponding to the applied current are shown in the insets. The dashed line is a guide to the eye.
When comparing the two I-V-curves they agree well within errors. The critical current for the onset of flux flow is slightly smaller during the neutron experiment ($I_{c_{exp}}\approx 30\,\text{A}$ vs $I_{c_{pre}} \approx 35\,\text{A}$). After flux flow is established the data points match well.
The slight deviations can be explained by the different time scales on which the I-V-curves were recorded, as mentioned above.

\section{Flux Line Bending}
\label{app:Fluxbending}
\begin{figure*}
	\centering
	\includegraphics[width = \textwidth]{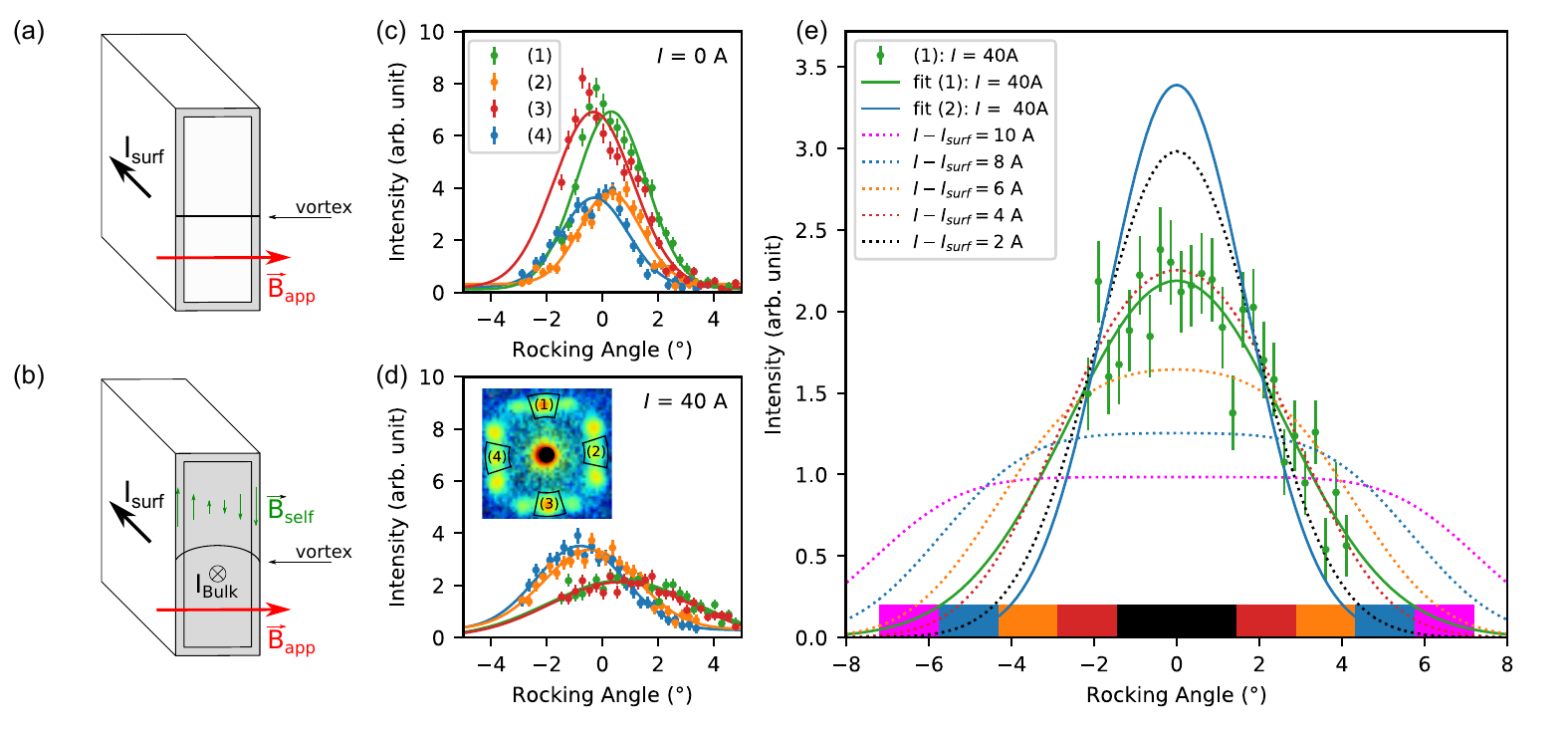}
	\caption{(a) Scheme of a straight flux line due to the absence of a bulk current. (b) Scheme of a bent flux line caused by the self-field contribution of a non-vanishing bulk current. (c) Rocking curves and fits (solid lines) of the first order Bragg peaks around a horizontal axis $\phi$ [curve (1) and (3)] and a vertical axis $\omega$ [curve (2) and (4)] with no applied current. (d) Rocking curves and fits (solid lines) of the first order Bragg peaks around a horizontal axis $\phi$ [curve (1) and (3)] and a vertical axis $\omega$ [curve (2) and (4)] with an applied current of $I = 40$\,A. The inset shows the sectors used for the rocking curves in panel (c) and (d). (e) Simulated rocking curves (dotted lines) for different combinations of bulk and surface current. The corresponding angular distribution is shown in the same color code at the bottom of the plot. }
	\label{fig:sup2}
\end{figure*}

We see from Ampere's law [Eq. (\ref{eq:maxwell})] that a transport current with a non-vanishing bulk current $I_{bulk}$ leads to curved flux lines due to its self field $B_{self}$, whereas a pure surface current $I_{surf}$, retains the alignment parallel to the external magnetic field $B_{app}$ as schematically depicted in Figs. \ref{fig:sup2} (a,b).
\begin{equation}
\label{eq:maxwell}
\mathbf{\nabla} \times \mathbf{B} = \mu_{0} \mathbf{J}
\end{equation}

The curvature of the flux lines can be extracted from the width of rocking scans of the FLL Bragg peaks as previously reported \cite{Kroeger1976, Pautrat2003}. The shape of the flux lines is affected by the self field resulting from the current distribution of a superconducting current-carrying wire. There are arguments for both, a constant, $r$ independent current distribution $J(r) = const$ \cite{Pautrat2003} and a square-root dependent current distribution $J(r) \propto \sqrt{1/r}$ \cite{Cubitt2009}, where $r$ is the distance from the sample's center.
Here we limit the discussion to an assumed homogeneous current distribution over the cross section of the Nb strip resulting in a self field, that linearly decreases when approaching the center of the sample. The resulting maximum tilt angles of the flux lines are given by $\theta = \pm B_{self} / B_{app}$ with the maximum value of the magnetic self-field $B_{self}$ on the surface of a strip with cross  section $w \times t$ with $w>t$ and current $I$, approximated by
\begin{equation}
\centering
B_{self} = \frac{\mu_{0} I}{2w}.
\label{eq:Bself}
\end{equation}
The resulting broadening of the rocking curve is given by

\begin{equation}
\centering
\text{FWHM}(\phi) = \Delta \phi = 2 \cdot \frac{\mu_{0} I}{2w B_{app}}.
\label{eq:phi_broadening}
\end{equation}
Due to geometry, the flux lines are only bent in the y-direction, which means only the rocking curves in $\phi$ around a horizontal axis are affected by the magnetic self-field broadening.

Figures \ref{fig:sup2} (c - f) give a detailed look on the rocking scans of the vortex lattice Bragg peaks in horizontal and vertical direction for different combinations of bulk and surface current. The sectors $1-4$ used for the rocking scans are marked in the inset of Fig. \ref{fig:sup2} (d).
Figures \ref{fig:sup2} (c - d) compare rocking curves with Gaussian fits (solid lines) of the first order Bragg peaks around a horizontal axis $\phi$ [curve (1) and (3)] and a vertical axis $\omega$ [curve (2) and (4)] for different applied currents $I$. We fitted a Gaussian with integrated intensity $I_{0}$, width $\sigma$, center $x0$ and y-offset $y_0$. The fit parameters of the fits are summarized in Table \ref{tab:sup1}.

We first describe the change in rocking curves in $\omega$ around a vertical axis [curves (2) and (4)] after a current of $I = 40\,$A is applied.
The rocking curves show an increase in integrated intensity $I_0$ by a factor of $\approx 1.2 -1.3$.  We only see a slight increase in rocking width [$\Delta \sigma \approx 0.5 - 0.6\,^{\circ}$]. We observe a shift of the respective rocking centers $x_0$ [$\Delta x_0 \approx  - (0.5 - 0.7)\,^{\circ}$].

The rocking curves in $\phi$ around a horizontal axis [curves (1) and (3)] show a decrease in integrated intensity $I_0$ by a factor of $\approx 1.5 - 1.8$ after a current of $I = 40\,$A is applied. We see a broadening of the rocking curve width $\sigma$  ($\Delta \sigma \approx 1.1 - 1.3^{\circ}$). In contrast to the $\omega$ rocking curve, we see a shift of $x_0$ in positive direction [$\Delta x_0 \approx  + (0.4 - 1.0)\,^{\circ}$].

Figure \ref{fig:sup2} (e)  shows simulated rocking curves (dotted lines) for different combinations of bulk and surface current and the rocking scan of Bragg spot (1) with applied current of $I = 40\,\text{A}$. The solid lines are the fits. All curves are scaled such that their integral is equivalent to the integral of the 40A fit.  The rocking curves were calculated  by approximating the magnetic self field according to Eq. (\ref{eq:Bself}) and convolving the resulting rectangular-shaped angular distribution with the fitted Gaussian of the 40\,A rocking curve of horizontal Bragg spot (2) \footnote{Using the width of the horizontal Bragg peak, which we assume is independent of magnetic self field broadening, includes the disorder due to the flux flow.}.  The corresponding angular distribution is shown in the same color code at the bottom of the plot. For low bulk currents, the shape of the calculated rocking curve is dominated by the Gaussian fit. For high bulk currents the shape is dominated by the rectangular angular distribution and we see the flat top shape as expected from a rocking curve corresponding to a continuously bent flux line.

When comparing the calculated rocking curves with the measured rocking curve in $\phi$ of Bragg peak  (1) at $I = 40\,\text{A}$,  the experimental curve is best approximated by the simulated curve with a bulk current contribution of $I_{bulk} = 4\,\text{A}$, which results in a surface current of $I_{surf} = 36\,\text{A}$.

\newcolumntype{C}[1]{>{\centering\arraybackslash}p{#1}}
\newcolumntype{R}[1]{>{\raggedleft\arraybackslash}p{#1}}

\begin{table}
	\caption{Fit parameters of Gaussian fits of the rocking scans around a horizontal and vertical axis of the Bragg peaks in vertical and horizontal direction for no applied current and a current of $I = 40\,\text{A}$. We fitted a Gaussian  with integrated intensity $I_{0}$, width $\sigma$, center $x0$ and y-offset $y_0$.}
	\label{tab:sup1}
	\centering
	\begin{tabular}{C{0.8cm}C{1cm}C{1.4cm}C{1.7cm}C{1.5cm}C{1.3cm}} 
		\hline\hline
		curve & current  & $I_{0}$ & $x_{0}$  & $\sigma$  & $y_{0}$  \\
		& (A)         &  (arb. unit) &  ($^\circ$) &  ($^\circ$) & (arb. unit) \\ [0.5ex] 
		
		\cline{1-6}
		\multicolumn{6}{l}{Rocking scans around horizontal axis (rocking angle $\phi$)}\\[0.2ex]
		\cline{1-6}
		(1) & 0  & $21.6 \pm 1.2$  & $0.32 \pm 0.05$ & $1.26 \pm 0.06$ & $0.1 \pm 0.1$ \\
		& 40 & $14.3 \pm 1.3$  & $0.67 \pm 0.16$ & $2.61 \pm 0.31$ & $0.0 \pm 0.4$ \\ [1ex]
		(3) & 0  & $23.9 \pm 1.7$  & $-0.32 \pm 0.09$ & $1.42 \pm 0.10$ & $0.2 \pm 0.1$ \\
		& 40 & $13.3 \pm 6.0 $ & $0.65 \pm 0.13$ & $2.50 \pm 0.55$ & $0.0 \pm 0.5$ \\
		\cline{1-6}
		\multicolumn{6}{l}{Rocking scans around vertical axis (rocking angle $\omega$)}\\[0.2ex]
		\cline{1-6}
		(4) & 0  & $10.3 \pm 1.0$  & $-0.28 \pm 0.06$ & $1.20 \pm 0.09$ & $0.2 \pm 0.2$ \\
		& 40 & $12.8 \pm 1.0$  & $-0.80 \pm 0.05$ & $1.59 \pm 0.09$ & $0.3 \pm 0.1$ \\ [1ex] 
		(2) & 0  & $9.7 \pm 0.8$   & $0.29 \pm 0.05$ & $1.10 \pm 0.07$ & $0.3 \pm 0.1$ \\
		& 40 & $12.9 \pm 2.0$  & $-0.41 \pm 0.06$ & $1.69 \pm 0.15$ & $0.3 \pm 0.2$ \\ [1ex]
		\hline\hline
	\end{tabular}
\end{table}

\section{Radial width of Bragg Peaks}
\label{app:rad_width}
\begin{figure}
	\centering
	\includegraphics[width = \columnwidth]{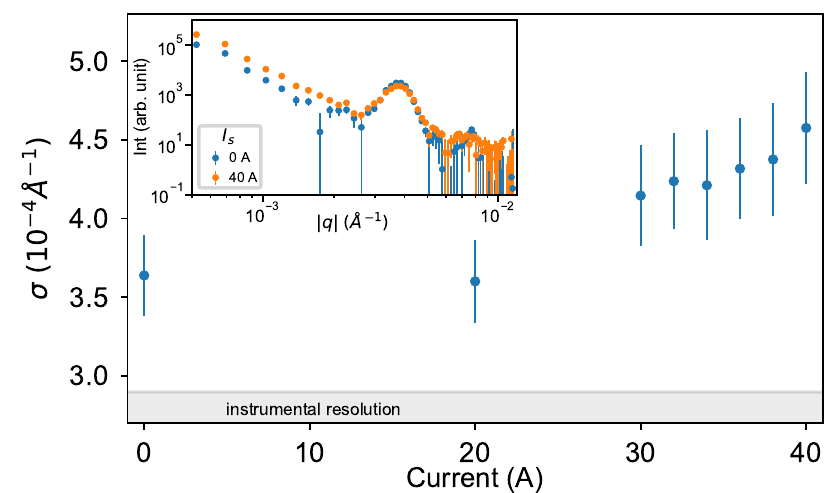}
	\caption{ Average radial width of all first order Bragg peaks as a function of current $I$ in a magnetic field of $B = 500$\,G at $T = 4$\,K. The inset shows the radial average of an exemplary single Bragg peak, here in horizontal direction, for $I = 0$\,A and $I = 40$\,A. The resolution limit of the instrument is colored in grey. }
	\label{fig:sup3}
\end{figure}

We can generate radial averages in $|\mathbf{q}|$ to extract the radial width $\sigma_{q}$ of our Bragg peaks. The radial width $\sigma_{q}$ is inversely related to the radial correlation length and is therefore a measure of the size of the well ordered mixed state domains.
Figure \ref{fig:sup3} shows the average radial width of all first order Bragg peaks from SANS measurements FC in a magnetic field of $B_{app} = 500\,\text{G}$ at $T = 4\,\text{K}$ under liquid He as a function of current $I$.  The inset shows the  radial average of an exemplary Bragg peak, here in horizontal direction, for $I = 0\,\text{A}$ and $I = 40\,\text{A}$. The resolution limit of the instrument is colored in grey. The radial width of each pair of Bragg spots at a given current $I$ was fitted individually. Within errors, there was no difference between the radial width of horizontal and vertical Bragg spots. Therefore the average radial width of all first order Bragg peaks is shown. We see a slight increase of the average radial width with increasing current starting at $I = 30\,\text{A}$. This relates to a decrease in radial correlation length and therefore to a shrinking of the well ordered mixed state domains, for both horizontal and vertical Bragg spots.
From the radial average of the horizontal Bragg peaks shown in the inset we can additionally clearly see the increasing intensity in horizontal direction at low q values for increasing current as also seen in the azimuthal averaging of the IMS scattering and the corresponding rocking scans (see Fig. \ref{fig:fig3} in the main text).

\section{Theoretical model}
\label{app:theory}

Theoretical modeling of the IMS dynamics is done using a two-component GL model \cite{da_silva_giant_2015} defined by the free energy density functional
\begin{align}
\label{eq:functional}
f =&\sum_{\nu=1,2}  \Big ( \frac{1}{2m_\nu} \big| {\bf D} \Psi_\nu \big|^2 + \alpha_\nu \big| \Psi_\nu \big|^2 + \frac{\beta_\nu }{2} |\Psi_\nu |^4 \Big)  \notag \\
& - \Gamma \big\{ \Psi_1^* \Psi_2 + \Psi_1  \Psi_2^* \big\}  +\frac{{\bf B}^2}{8\pi}, 
\end{align}
where $\Psi_{1,2}({\bf r})$ are the condensate components and ${\bf D} = - i \hbar  \nabla - 2e {\bf A}/c$. For a two-band system, the temperature-dependent coefficients $\alpha_\nu$, $\beta_\nu$,  $m_\nu$, and  the interband Josephson-like coupling constant $\Gamma$  are derived from the microscopical model for the band carrier states. Here, we apply this model to describe the IMS in a single-band superconductor making use of the qualitative similarities between the IT regime in single- and two-band materials.

Employing this two-component model circumvents a well-known difficulty of the theoretical description of the IMS of a IT superconductor. The GL theory is not applicable in this case, because according to it the IT domain between types I and II degenerates into a single critical point $\kappa = \kappa_0$, where all vortex configurations are degenerate. On the other hand, solving a full set of the microscopic equations is prohibitively expensive computationally, especially for irregular IMS vortex configurations.  Stationary IMS configurations have been recently reproduced using the extended GL theory \cite{vagov_universal_2020}. But the corresponding time-dependent version of this approach suitable to study the current-driven non-stationary IMS is not yet available. 

The two-component model offers a reasonable alternative that captures many properties of the IMS at least qualitatively. This model goes beyond the conventional GL theory which for a two-band system has still a single order parameter \cite{kogan_ginzburg-landau_2011,shanenko_extended_2011}. This fact is derived also from the two-component model where  in the limit $T\to T_c$ both components $\Psi_1$ and $\Psi_2$ have the same spatial profile. However, this model accounts for the non-local effects responsible for the appearance of the finite IT domain when $T< T_c$  \cite{da_silva_giant_2015}. A detailed comparison of the two-band \cite{da_silva_giant_2015} and single-band models \cite{wolf_vortex_2017,vagov_universal_2020} demonstrates that the key IMS features in both models are equivalent.

Equations of the two-component model are derived from the extremum condition for the energy functional (\ref{eq:functional}) and additional time derivatives are introduced to describe the dynamics. It is convenient to write these equations using a system of units defined by the zero-temperature coherence length, uniform solution for the gap and for the critical field, calculated for one of the equations separately \cite{da_silva_giant_2015}. This yields
\begin{subequations}
	\begin{align}
	\label{eq2a}
	&\eta D_t  \psi_1=
	{\bf D}^2 \psi_1-(\chi_1-|\psi_1|^2)\psi_1 - \gamma \psi_2, \\
	\label{eq2b}
	&\eta  D_t  \psi_2 =
	\frac{1}{\alpha}  {\bf D}^2 \psi_2-(\chi_2-|\psi_2|^2)\psi_2 - \frac{\beta_2}{\beta_1} \gamma \psi_1, \\
	&\kappa_1^2 \,\boldsymbol{ \nabla} \times \boldsymbol{ \nabla} \times {\bf A} =\frac{1}{\kappa_1^2}\Re\big[ \psi_1 {\bf D} \psi^*_1 \big] + \frac{\alpha}{\kappa^2_2} \Re \big[ \psi_2 {\bf D} \psi_2^* \big],
	\end{align}
	\label{eq:GL_equations}
\end{subequations}
where ${\bf D} = - i \nabla - {\bf A}$ and $D_t = \partial_ t -i\phi$ are the scaled gauge-invariant derivatives with ${\bf A}$ and $\phi$ being the vector and scalar potentials, respectively, $\gamma$ denotes the (scaled) interband coupling constant, $\kappa_{1,2}$ are the GL parameters for the components $\Psi_{1,2}$, calculated for each of the component equations (\ref{eq2a}) and (\ref{eq2b}) separately at $\Gamma = 0$,   and $\alpha = \xi_2^2/\xi_1^2$ where $\xi_{1,2}$ are the GL coherence lengths calculated separately for both components. The GL parameters are related as $\kappa_2 = \alpha \kappa_1 \sqrt{\beta_1/\beta_2}$. Notice, that the GL parameter $\kappa$  for the entire system (at $T \to T_c$) differs from both $\kappa_1$ and $\kappa_2$, its expression can be found in Ref. \cite{kogan_ginzburg-landau_2011}.  Parameters $\chi_{\nu} = \tau - S_\nu$ with $\tau = 1 - T/T_c$ define the temperature dependence. Here constants $S_\nu$ are determined by the intraband coupling. They satisfy the condition $S_1 S_2  \alpha^2 = \gamma^2 \beta_1/\beta_2$ which ensures that the superconducting order parameter disappears at $\tau \to 0$, so that $T_c$ is the superconductivity transition temperature (one can find a detailed derivation of all parameters for a two-band model in Ref. \cite{da_silva_giant_2015}). Finally,  $\eta$ is related to the system losses and determines the time scale. The influence of the normal current on the evolution of the IMS configurations is neglected. 

In the calculations we take $S_1 = 0.043$, $S_2 = 0.188$, $\beta_1/\beta_2 = 0.92$,  $\kappa_1 =1.5$. We also assume  $\alpha^2 = 0.65$ and $T=0.74T_c$ which ensures that the system is in the IT domain, very close to the line of zero surface tension of the N-S domain wall, where the IT superconductivity is expected. These values are taken from the earlier work, where the equivalence of the two- and one-component models was established   \cite{da_silva_giant_2015}. However, qualitative features of the IMS dynamics remain qualitatively similar in a wide range of parameters as long as the system is in the IT regime. 

\begin{figure*}
	\centering
	\includegraphics[width = 1.8\columnwidth]{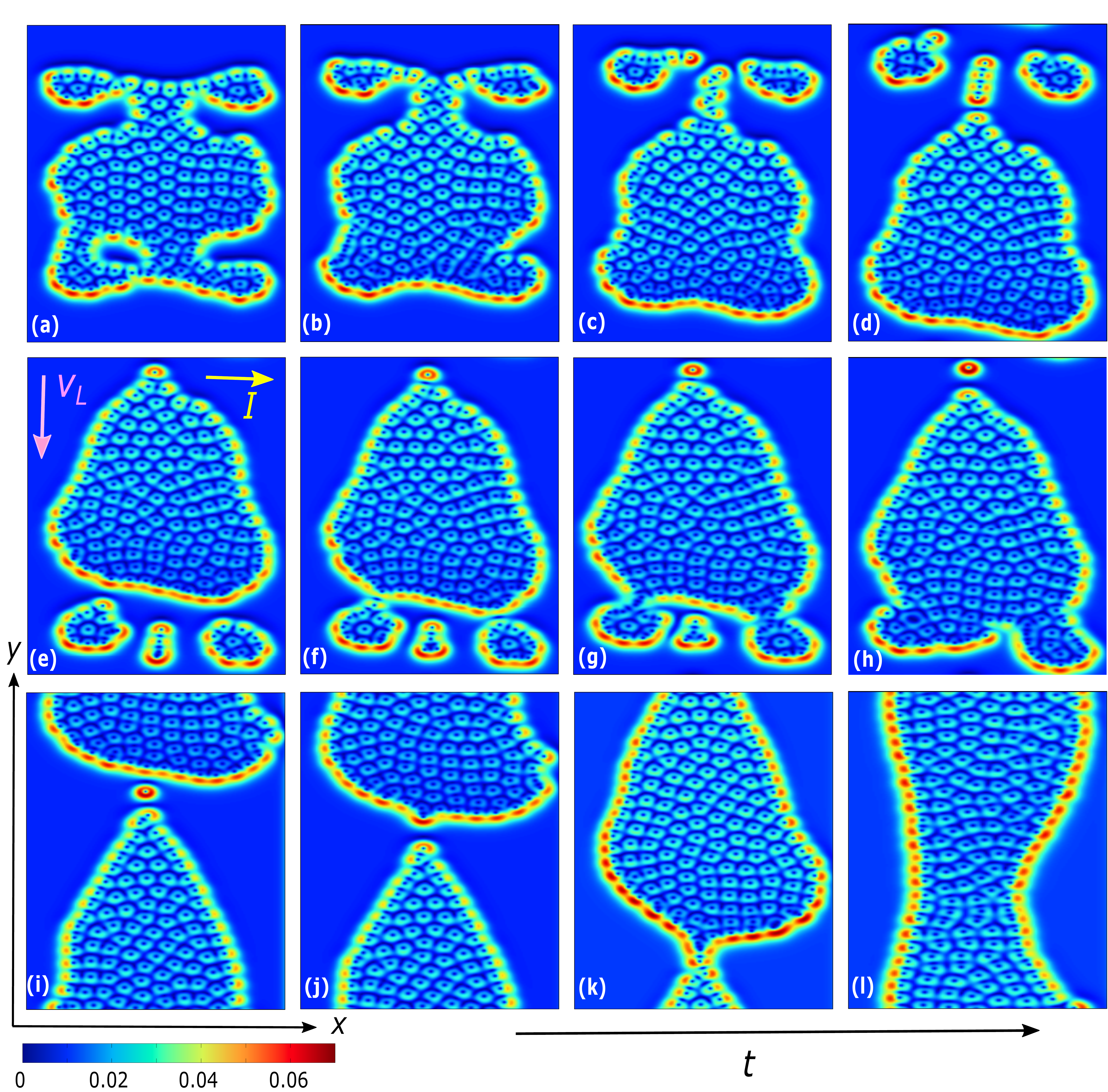}
	\caption{The spatial dependence of the current  density calculated for a vortex cluster at chosen time instants illustrating the time evolution (the time direction is shown by an arrow marked $t$), panel (a) is the initial state and panel (l) is the final.  Current $I$ flows horizontally rightwards, vortices move vertically downwards.  In panels (a - h) the observation frame shifts downwards to keep the largest cluster in focus to demonstrate the time evolution of its shape and relative positions of neighboring smaller clusters. Areas of the larger current (red) correspond to a larger pulling force acting normal to the cluster boundary. Each panel row illustrates a different stage of the evolution, which produces vertical vortex stripes in the end.
	}
	\label{fig:sup4}
\end{figure*} 

We consider a geometry where vortices are directed along the $z$ axis and the corresponding magnetic field is then ${\bf B} = (0,0,B(x,y))$. We also assume that in the $y$ direction  the sample has the finite width $a$ and that the current flows between the sample surfaces $x=0,a$. The length in the $y$ direction is taken much larger, $L \gg a$.  A complete description of the current-induced vortex evolution requires solving Eqs. (\ref{eq:GL_equations}) for a realistic finite-size sample together with the equation for the charge density distribution in the sample as well as in the contacts. In this work we follow a simplified approach where the current is "injected" into the superconductor by applying a difference of the phase/field at the sample boundaries. Notice, that deep inside a sample the current is absent in the Meissner phase and therefore isolated vortex clusters do not move.  However, a current flowing at the surface interacts with vortices creating the Lorentz force [Fig. \ref{fig:sup2} (b)]. We model this situation by injecting a small current by imposing the linear potential  $\phi(x,y)=g x$ which creates a phase difference between the boundaries $x=0$ and $x=L$, where $g$ controls the current value. This model neglects the current decay inside the sample and thus does not describe the vortex bending, discussed in the main text. It is nevertheless sufficient to capture general features of the evolution of the IMS superstructures.  

Equations (\ref{eq:GL_equations}) are solved with the  superconductor-metal boundary conditions ${\bf D}_\perp \psi_j=-i \psi/b$ applied at $x=0$ and $x=L$. A particular value of the real parameter $b$ is not important, it is set to $b = 20$ in the calculations. In the $x$ direction we assume the periodic boundary conditions. We note, that details of the model geometry and the boundary conditions have only little influence on the  dynamics of the vortex matter far from the sample boundaries. Equations (\ref{eq:GL_equations}) are solved on a two-dimensional grid with the spacing $a_x  =  a_y  = 0.25 \xi_1$, which is sufficient to describe vortices. In the calculations we first obtain a stationary vortex configuration by solving Eqs. (\ref{eq:GL_equations}) with $g=0$. Then, the potential is switched on ($g \neq 0)$, and the evolution of the IMS begins. We allow the vortex matter to evolve until it  achieves a quasi-stationary configuration (stripes). 

\section{Details of the vortex cluster evolution}
\label{app:elongation}

Results of the calculations are shown in Fig. \ref{fig:fig4} in the main text and in Fig. \ref{fig:sup4}. When an IT superconductor is placed in the magnetic field and the current is absent, vortices inside the sample form the IMS of randomized vortex clusters [Fig. \ref{fig:fig4} (a)].  When the current flows the IMS changes, eventually forming stripes in the direction perpendicular to the current. We note that the evolution is qualitatively similar for all initial IMS configurations. Figures \ref{fig:sup4} (a)-(l) illustrate main stages of the evolution by showing snapshots of the current density for a relatively large isolated vortex cluster, where each row of the figure highlights a specific feature in the elongation process. To demonstrate the time evolution of the clusters shape, the observation frame in Figs. \ref{fig:fig4} (a - h) moves downward and keep the largest cluster in focus so that one observes the relative motion of different clusters, all moving downwards [in the last row of Figs. \ref{fig:fig4} (i - l) the frame does not move].

By looking at Fig. \ref{fig:sup4} we first note that the current distribution in voids of the Meissner phase inside the cluster forces them to shrink and disappear [cf.  Figs. \ref{fig:sup4} (a) and (b)]. The second important feature to note is that the total current-induced force increases with the boundary length such that larger clusters (or separate parts of a cluster) move faster.  This is clearly seen in Figs. \ref{fig:sup4} (a) -  (d) where smaller semi-isolated structures at the rear move slower than the main cluster part and  break off eventually. The evolution shown in Figs. \ref{fig:sup4} (e - h) illustrates the same trend from the opposite perspective: here a large fast-moving cluster catches up with and then absorbs the smaller ones. 

While moving, the cluster gradually elongates in the vertical direction which takes place due to a combined action of the Lorentz force, that pulls the cluster on its lower boundary, and the drag force, that slows down vortices at its rear. As a result, the cluster acquires the shape similar to that of a liquid droplet falling in the air. Vortices at the rear part of the cluster are only weakly coupled to its main body and are eventually broken off and left behind. This process makes the neighboring clusters merge, forming a stripe in the direction of their motion [Fig. \ref{fig:sup4} (i)-(l)]. 

The evolution of vortex clusters can be intuitively understood as a combination of three main factors: the Lorentz force, the drag force and the inter-vortex interaction potential. These are typical for the  superconducting mixed state, however, their action on the IMS in IT superconductors has a number of specific features. In a conventional type II superconductor the distance between vortex cores is much larger then their size so that each vortex can be considered separately with respect to the acting Lorentz and drag forces. Therefore the current-induced dynamics is practically the same for all vortices (although even for type II superconductors the vortex motion changes the inter-vortex interaction making it dependent on the vortex mutual position \cite{kogan_2020}). 

In contrast, the size of the vortex core in IT superconductors is comparable to the inter-vortex distance determined by the minimum of the non-monotonic vortex-vortex interaction potential \cite{vagov_superconductivity_2016}. In addition, multi-vortex interactions play an increasingly important role for large vortex clusters \cite{da_silva_giant_2015}. As a consequence, vortices can no longer be regarded as separate "elementary particles" of the mixed state. The properties of a vortex cluster therefore do not depend simply on its number of vortices but also on the cluster configuration – shape and size. The applied current is strongly distorted by the vortex configuration flowing mainly along cluster boundaries (see Fig. \ref{fig:sup4}). Thus the Lorentz force acts mainly on the boundary pulling it in the normal direction. In addition the current profile is notably asymmetric, so that its density is larger at the down-facing boundaries and consequently leads to a larger net force in the downward direction. The drag force acting on a vortex inside a cluster is smaller than that for a separate vortex because the movement inside a cluster involves smaller changes in the field-condensate profile due to the comparable size of the vortex core to the intervortex distance. Then the drag force is largest for the boundary vortices (we note that our model does not take into account all mechanisms leading to the vortex drag \cite{kopnin_1976,blatter_vortices_1994}, however, this does not change qualitative conclusions). The dependence of the Lorentz and drag forces on the cluster configuration and size gives rise to the velocity dispersion for different clusters and cluster parts that is clearly visible in the numerical calculations in Fig. \ref{fig:sup4}. 

Finally, the unique vortex interactions in an IT material ensure the preferred mean intervortex distance, but not the vortex cluster shape. The Lorentz force, that pulls the cluster boundary downwards, the drag, that acts on vortices in the opposite direction and a "soft" inter-vortex interaction result in elongated clusters in the direction of the movement, i.e. perpendicular to the current. The elongation increases the bypassing current flow and thus the Lorentz force at the boundary which, in turn, leads to a still faster elongation.


%

\end{document}